\documentclass[aps,prl,twocolumn,superscriptaddress,showpacs]{revtex4}

\usepackage{amsmath}
\usepackage{amssymb}
\usepackage{amsfonts}
\usepackage{mathrsfs}
\usepackage{graphicx}
\usepackage[british]{babel}

\begin{document}

\title{The advantage of L{\'e}vy strategies in intermittent search processes}

\author{Michael A. Lomholt}
\affiliation{Physics Department, Technical University of Munich,
D-85747 Garching, Germany}
\author{Tal Koren}
\affiliation{School of Chemistry, Tel Aviv University, 69978 Tel Aviv,
Israel}
\author{Ralf Metzler}
\affiliation{Physics Department, Technical University of Munich,
D-85747 Garching, Germany}
\author{Joseph Klafter}
\affiliation{School of Chemistry, Tel Aviv University, 69978 Tel Aviv,
Israel}

\begin{abstract}
Search strategies based on random walk processes with long-tailed jump length
distributions (L{\'e}vy walks) on the one hand and intermittent behavior
switching between local search and ballistic relocation phases on the other,
have been previously shown to be beneficial in stochastic target finding
problems. We here study a combination of both mechanisms: an intermittent
process with L{\'e}vy distributed relocations. We demonstrate how L{\'e}vy
distributed relocations reduce oversampling and thus further optimize the
intermittent search strategy in the critical situation of rare targets.
\end{abstract}


\maketitle

Random search processes occur in many areas. The simplest example is that of
passive particles immersed in a thermal bath subjecting them to Brownian motion
until encounter, for instance, in chemical reactions \cite{smoluchowski}. This
Brownian search dynamics may be accelerated in various ways: (i) By a drift
toward the reaction center, for instance, in the time-dependent Onsager problem
for diffusion in an attractive Coulomb potential,
or in chemotaxis of biological cells \cite{hong,chemo}.
(ii) By combining more than one search mechanism due to available interfaces as
known from gene regulation \cite{bvh}: to find their target sequence on the
DNA molecule more efficiently, proteins switch between 3D bulk diffusion, and
1D sliding along the DNA \cite{igor}. (iii) By performing a L{\'e}vy walk, i.e.,
a random walk whose jump lengths are drawn from a long-tailed distribution
$\lambda(x)\simeq |x|^{-1-\alpha}$ ($0<\alpha<2$) \cite{klablushle}. The
resulting trajectory has diverging variance $\langle x^2\rangle=\infty$, unless
a velocity is introduced, and fractal dimension $d_f=\alpha$,
covering space less densely to reduce oversampling, an advantage over
Brownian search \cite{klafter86,viswanathan99,michael,dartel04,reynolds}. (iv)
By intermittent strategies during which local (Brownian) search switches with
ballistic relocations \cite{benichou05}.

We here demonstrate for a searcher without orientational memory how
intermittent and L{\'e}vy search strategies can be
combined to produce a synergistic strategy, that for rare targets is more
efficient than previously introduced intermittent search models. Similarly
to Refs.~\cite{benichou05,bartumeus02,benichou06,oshanin07}
we focus on the 1D case, that is
relevant, for instance, for animals searching for food at ecological
interfaces (forest edges, coastlines etc.).

Generalizing the search model from
Ref.~\cite{benichou06}, we consider two phases: In phase 1 the searcher looks
for
the target performing diffusive motion with diffusion constant $D$. There is a
probability per time $\tau_1^{-1}$ that the searcher leaves this search phase
and switches to phase 2, the relocation phase, where it moves ballistically
with velocity $v$ in a random direction. The time spent relocating is drawn
from the waiting time distribution $\psi(t)$, that previously was taken to be
exponential (leading to a Markovian process) \cite{benichou05,benichou06}, but
we relax this assumption here to show the advantage of L{\'e}vy strategies.
The purpose of the relocation phase is to move as quickly as possible away from
the area that has just been searched, and thus the searcher is not scanning
for the target in this phase. To compare with previous results
we take a closed cell approach: the
search is performed on an interval of length $L$ with periodic boundary
conditions, corresponding to regularly spaced targets with density $1/L$. The
model can be formulated as an equation for the probability density $P(x,t)$
for the position $x$ of the searcher in the search phase:
\begin{eqnarray}
\frac{\partial P}{\partial t}&=&\frac{1}{\tau_1}\int_{-L/2}^{L/2}d x'\int_0^t d
t'\;W(x-x',t-t')P(x',t')\nonumber\\ &&-\frac{1}{\tau_1}P(x,t)+D\frac{\partial^2
P}{\partial x^2}-p_{\rm fa}(t)\delta(x).
\label{eq:master}
\end{eqnarray}
The role of the last term on the right hand side is to remove the particle
when it arrives at the target placed at $x=0$. The density $p_{\rm fa}(t)$
thus represents the first arrival time at the target, which
is determined implicitly by the absorbing boundary condition
$P(x=0,t)=0$. The kernel $W(x,t)$ representing relocations is given by
\begin{equation}
W(x,t)=\frac{\psi(t)}{2}\sum_{n=-\infty}^{\infty}\delta(|x+n L|-v t)\; .
\end{equation}
The $\delta$-coupling enforces that the distance traveled in time $t$ is $vt$,
and the sum over $n$ renders $W(x,t)$ $L$-periodic in $x$. $\psi(t)$ is related
to the spatial distribution of the relocations $\lambda(x)$ by $\psi(t)=2 v
\lambda(v t)$. $\lambda(x)$ is assumed to be symmetric around $x=0$ (no
orientational memory).

\begin{figure}
\includegraphics[width=8cm]{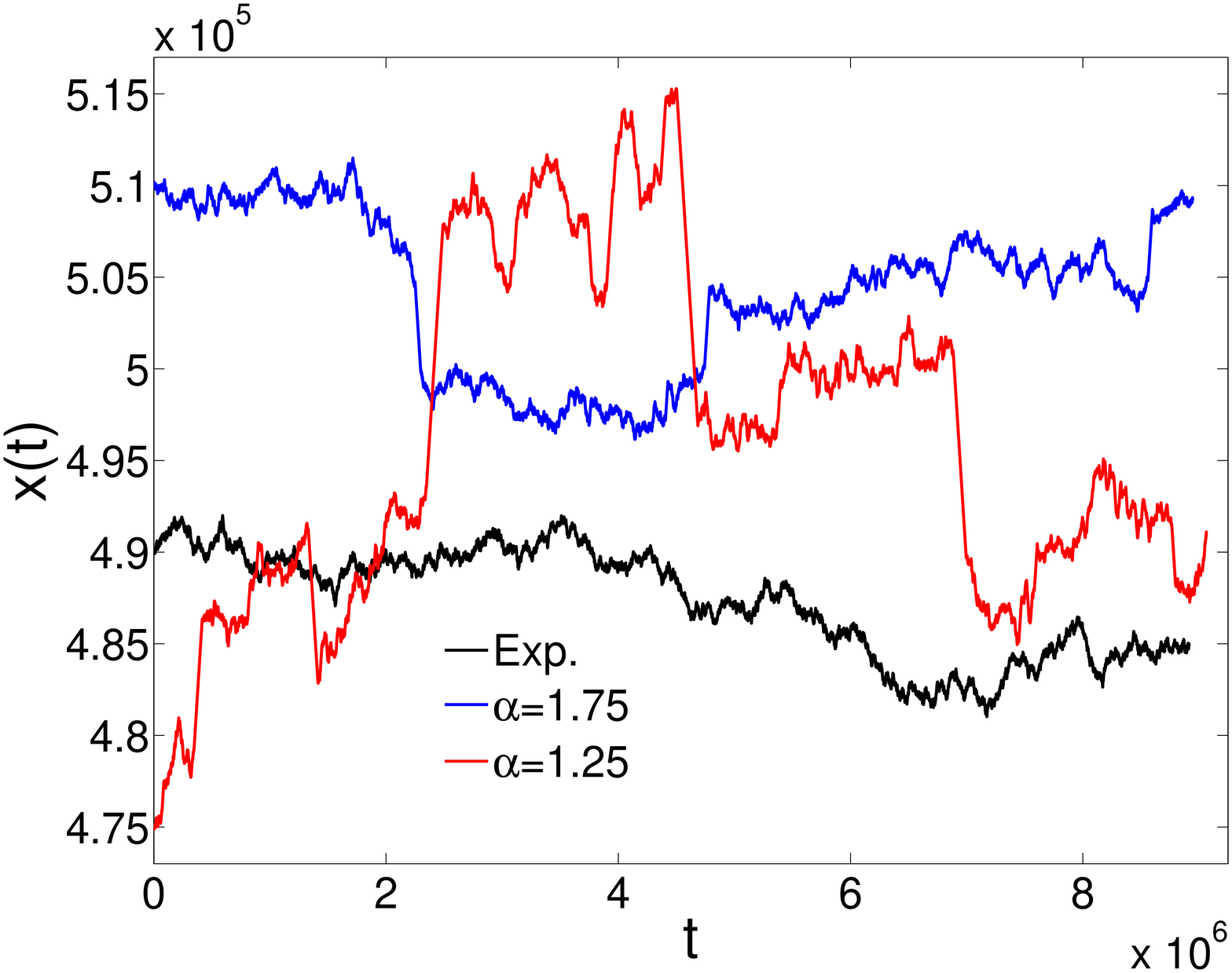}
\caption{$x$-$t$ diagram with exponential and L{\'e}vy relocations,
with $\tau_1=37$, $\tau_2=200$, $D=1$, $v=0.1$ and $L=\infty$.}
\label{fig1}
\end{figure}

The search efficiency is quantified by the mean search time $\left<t\right>=
\int_0^\infty d t\,t p_{\rm fa}(t)$. To obtain $\left<t\right>$ we Fourier
expand $P(n,t)=\int_{-L/2}^{L/2} d x\,e^{i k_n x}P(x,t)$ ($n$ is an integer
with corresponding wavenumber $k_n=2\pi n/L$), and Laplace transform $P(n,u)=
\int_0^\infty d t\,e^{-u t}P(n,t)$, to find
\begin{eqnarray}
u P(n,u)-\delta_{n,0}&=&\frac{1}{\tau_1} W(n,u)P(n,u)-\frac{1}{\tau_1}P(n,u)\nonumber\\
&&-D k_n^2 P(n,u)-p_{\rm fa}(u).
\end{eqnarray}
The initial distribution is uniform, $P(x,t=0)=1/L$, since the searcher
initially has no information on the position of the target. Isolating $P(n,u)$,
summing over $n$ (note that $\sum_{n=-\infty}^\infty P(n,u)=P(x=0,u)=0$), and
solving for $p_{\rm fa}(u)$ we find
\begin{equation}
p_{\rm fa}(u)=\left\{\sum_{n=-\infty}^\infty\frac{u+[1-\psi(u)]/\tau_1}{u+D k_n^2+[1-W(n,u)]/\tau_1}\right\}^{-1}\;.\label{eq:pfa}
\end{equation}
In Laplace space the mean search time $\left<t\right>$ can be found by
expanding $p_{\rm fa}$ at small $u$ since
$p_{\rm fa}(u)\sim 1-\left<t\right>u+\dots$.
Be $\tau_2$ the average time spent in one relocation event we have $\psi(u)
\sim1-\tau_2 u+\dots$, and thus arrive at
\begin{equation}
\left<t\right>=\sum_{n=1}^{\infty}\frac{2(\tau_1+\tau_2)}{D\tau_1k_n^2+1-
\lambda(k_n)}.
\label{eq:Tsearch}
\end{equation}
Here $\lambda(k_n)=W(n,u=0)=\int_{-\infty}^\infty d x\;e^{i k_n x}\lambda(x)$
is the Fourier transform of the
relocation length distribution
at the discrete wavenumbers
$k_n=2\pi n/L$. We now use Eq.~(\ref{eq:Tsearch}) to determine the search efficiency of
(i) L{\'e}vy and (ii) exponentially distributed relocations:

(i) For L{\'e}vy distributed relocations we use the symmetric L{\'e}vy stable
law with characteristic function \cite{metzler00}
\begin{equation}
\lambda(k)=e^{-\sigma^\alpha|k|^\alpha}\;,\quad \sigma=\frac{\pi v \tau_2}{
2\Gamma(1-1/\alpha)}.
\end{equation}
The index $\alpha$ is restricted to $1<\alpha<2$ so that the
mean relocation time $\tau_2$ is finite. Fig.~\ref{fig1} depicts
trajectories for cases of exponential and L{\'e}vy relocations, distinguishing
the L{\'e}vy case with its occasional long relocations.

We introduce three approximations valid for large $L$:

(a) Assume that $v\tau_2\gg\sqrt{D\tau_1}$, i.e., that the mean relocation
distance is much longer than the average distance scanned in a typical search
phase. We will see that this is self-consistent with the obtained optimal
values of $\tau_1$ and $\tau_2$ that have the same $L$-scaling for large $L$.
This assumption means that $D\tau_1 k_n^2$ and $\lambda(k_n)$ are to a good
approximation non-zero at different $n$, and we expand
\begin{equation}
\frac{1}{D\tau_1 k_n^2+1-\lambda(k_n)}
\sim\frac{1}{D\tau_1k_n^2+1}+\frac{1}{1-\lambda(k_n)}-1\;.\label{eq:approx1}
\end{equation}

(b) Assuming that the search range $\sqrt{D\tau_1}$ is much smaller than $L$,
we replace the sum over the first term on the right hand side of
Eq. (\ref{eq:approx1}) by an integral, yielding
\begin{equation}
\sum_{n=1}^{\infty}\frac{1}{D\tau_1 k_n^2+1}\sim\int_0^\infty\frac{1}{D\tau_1
k_n^2+1}dn=\frac{L}{4\sqrt{D \tau_1}}.
\end{equation}

(c) Approximate the last two terms of Eq.~(\ref{eq:approx1}):
as the contribution from the singularity at small $n$
dominates the sum (note that $k_n|_{n=1}\to 0$ in the limit of large $L$),
\begin{equation}
\sum_{n=1}^{\infty}\left[\frac{1}{1-\lambda(k_n)}-1\right]
\sim\left(\frac{L}{2 \pi \sigma}\right)^\alpha \zeta(\alpha).
\end{equation}
Here $\zeta(\alpha)=\sum_{n=1}^\infty n^{-\alpha}$ is the Riemann $\zeta$
function.

Collecting (a) to (c), Eq. (\ref{eq:Tsearch}) is approximated by
\begin{equation}
\label{eq:Tlevy}
\left<t\right>\sim 2(\tau_1+\tau_2)\left[\frac{L}{4\sqrt{D \tau_1}}+\left(
\frac{L}{2\pi\sigma}\right)^\alpha \zeta(\alpha)\right].
\end{equation}

\begin{figure}
\includegraphics[width=8cm]{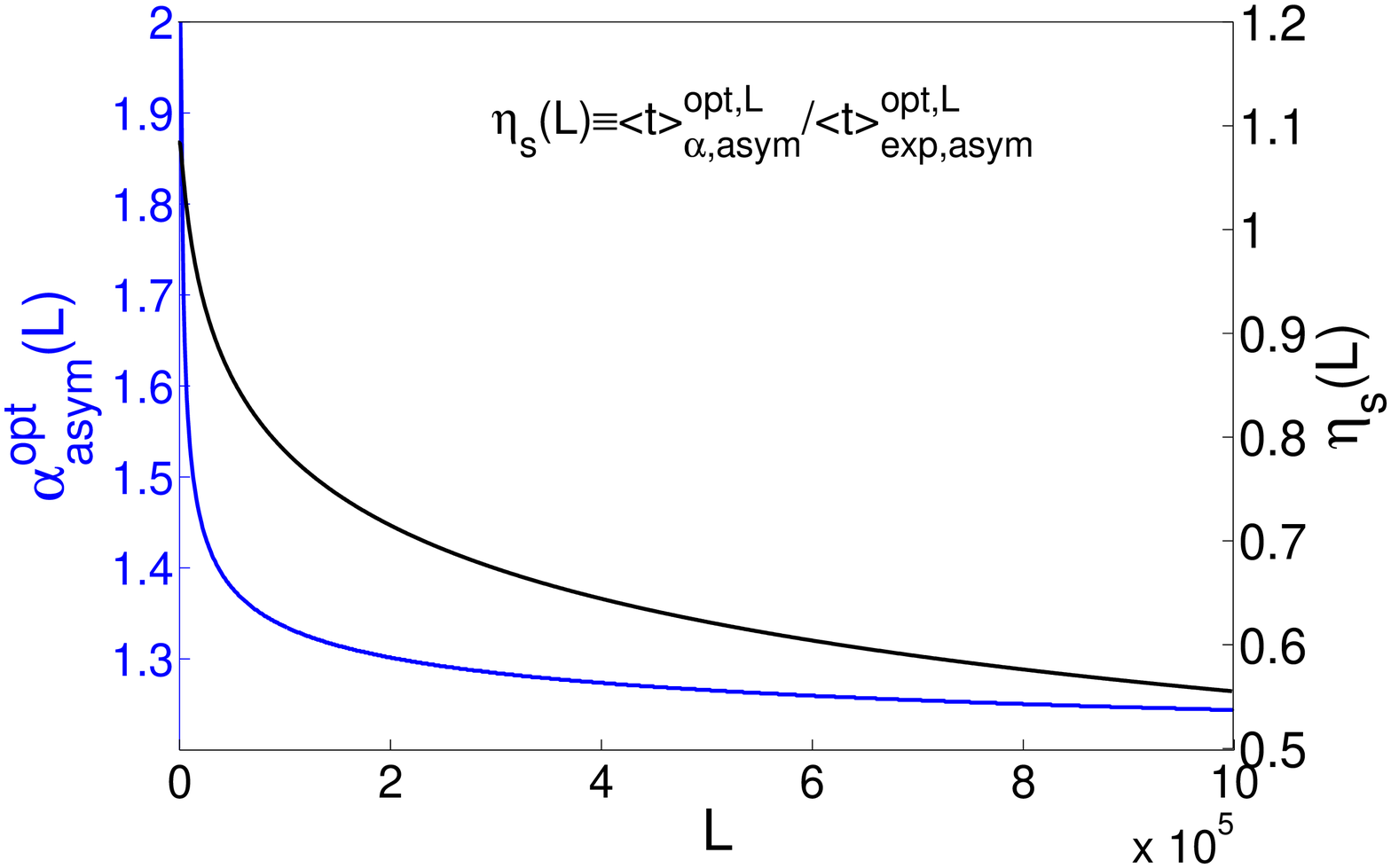}
\caption{Optimal $\alpha$, and ratio $\eta_s$ of search times for optimal
$\alpha$ versus exponential strategy, as function of $L$ ($D=1$ and $v=1$).
All values are calculated using the asymptotic Eqs. (\ref{eq:Tlevy}) and
(\ref{eq:Texp}), and corresponding optimal $\tau_1$ and $\tau_2$.}
\label{figopt}
\end{figure}

For honest comparison between L{\'e}vy and exponential strategies, we determine
the respective optimal $\tau_1$ and $\tau_2$. Solving $\partial\langle t\rangle/
\partial \tau_1=0$ and $\partial\langle t\rangle/\partial\tau_2=0$
simultaneously, we obtain from Eq.~(\ref{eq:Tlevy}) that at large $L$
\begin{equation}
\tau_1\sim (b/a^\alpha)^{1/(\alpha-1/2)}\;,\quad\tau_2\sim(b/\sqrt{a})^{1/
(\alpha-1/2)},
\label{eq:opttaulevy}
\end{equation}
where (using $\Omega\equiv\sqrt{1+4(\alpha-1)\alpha}$)
\begin{subequations}
\begin{eqnarray}
a=(1+\Omega)/(2[\alpha-1]),\hspace*{4cm}\\
b=2\sqrt{D}\left[2\alpha+\Omega-3\right]\zeta(\alpha)L^{\alpha-1}
\left[\frac{\Gamma\left(1-\alpha^{-1}\right)}{\pi^2 v}\right]^\alpha
\end{eqnarray}
\end{subequations}
such that the optimal $\tau_i$ scale with $L$ like $L^{(\alpha-1)/(\alpha
-1/2)}$. According to Eq.~(\ref{eq:Tlevy}), $\left<t\right>$ will then
scale like $L^{(3\alpha-2)/(2\alpha-1)}$, implying that for large $L$ the
more efficient search will occur for $\alpha$ close to 1. However, the
prefactor to the $L$-scaling diverges as $\alpha\to 1$, so the optimal choice
of $\alpha$ will be somewhat larger than 1 for any finite $L$, as demonstrated
in Fig.~\ref{figopt}.

(ii) For exponentially distributed relocation with
\begin{equation}
\psi(t)=\tau_2^{-1}e^{-t/\tau_2},
\end{equation}
approximations (a) to (c) also apply, with $\sigma=v\tau_2$.
The corresponding results for $\langle t\rangle$ and optimal $\tau_i$ obtain
by replacing $\Gamma(1-1/\alpha)$ by $\pi/2$ and taking $\alpha=2$:
\begin{eqnarray}
\left<t\right>\sim \frac{\tau_1+\tau_2}{12}\left[\frac{6 L}{\sqrt{D\tau_1}}+
\left(\frac{L}{v\tau_2}\right)^2\right],
\label{eq:Texp}\\
\tau_1\sim\frac{1}{2}\left(\frac{D}{18 v^4}\right)^{1/3}L^{2/3},\qquad
\tau_2\sim2\tau_1.
\label{eq:opttauexp}
\end{eqnarray}
These expressions agree with those of Ref.~\cite{benichou06,REM}.

\begin{figure}
\includegraphics[width=8cm]{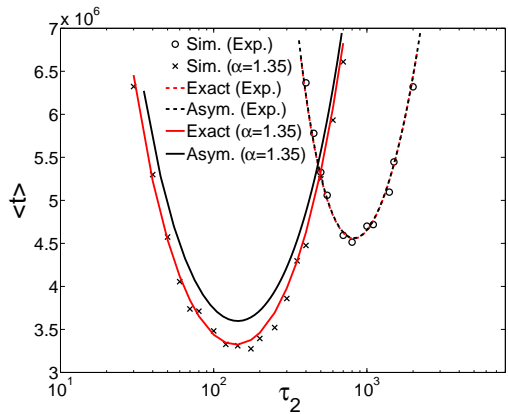}
\caption{Mean search time for L{\'e}vy ($\alpha=1.35$) and exponential
strategies as function of $\tau_2$ at asymptotically optimal $\tau_1$
($\tau_1=37.2$ for L{\'e}vy and $\tau_1=411$ for exponential). We chose
$L=10^5$, $D=1$, $v=1$. Simulations versus exact (Eq.~(\ref{eq:Tsearch}))
and asymptotic (Eqs.~(\ref{eq:Tlevy}) and (\ref{eq:Texp})) theory.}
\label{figcmp1}
\end{figure}

The search time $\left<t\right>$ with $L$ for exponential strategies scales
like $L^{4/3}$ for optimal $\tau_1$ and $\tau_2$. This proves that L{\'e}vy
strategies with $1<\alpha<2$ are increasingly more efficient than the
exponential strategies for decreasing target density. In Fig.~\ref{figcmp1}
we show $\langle t\rangle$ as function of relocation time $\tau_2$.

To understand better the
$\alpha$-dependence of the L{\'e}vy strategy we study the first arrival density
$p_{\rm fa}(t)$ for large $L$, where again $L\gg v\tau_2\gg \sqrt{D\tau_1}$. We
consider times much longer than one relocation-search cycle such that $\psi(u)
\sim 1-\tau_2 u+\dots$, and rewrite Eq.~(\ref{eq:pfa}) as
\begin{equation}
p_{\rm fa}(u)\sim\frac{1}{u}\frac{\tau_1}{\tau_1+\tau_2}\frac{1}{W_0(u)}\frac{1}{L}\;,\label{eq:pfaW0}
\end{equation}
where we have introduced the term
\begin{equation}
W_0(u)=\frac{1}{L}\sum_{n=-\infty}^\infty\frac{1}{u+D k_n^2+[1-W(n,u)]/\tau_1}\;.\label{eq:W0u}
\end{equation}
The last expression can be simplified following similar approximations as for
$\langle t\rangle$ before. The separation of length scales leading to
approximation (a) allows us to write
\begin{equation}
W_0(u)\sim \frac{\tau_1}{L}\sum_{n=-\infty}^\infty\bigg[\frac{1}{D\tau_1
k_n^2+1}
+\frac{1}{\tau_1 u+1-W(n,u)}-1\bigg].
\label{eq:W0ubefint}
\end{equation}
For the last two terms in Eq.~(\ref{eq:W0ubefint}) the contribution at small
$n$
again dominates the sum (approximation (c)), and we expand $W(n,u)$ at small
$k_n$ and $u$, finding $W(n,u)\sim 1-\sigma^\alpha|k_n|^\alpha-\tau_2 u$.
Collecting the results, we see that
\begin{equation}
W_0(u)\sim \frac{\tau_1}{L}\sum_{n=-\infty}^\infty\left[\frac{1}{D\tau_1
k_n^2+1}+\frac{1}{(\tau_1+\tau_2) u+\sigma^\alpha|k_n|^\alpha}\right].
\label{eq:W0ubefint2}
\end{equation}
We focus on times short enough such that the $L$-periodicity of the problem
does not yet play a role, so that Laplace space $u\gg (\sigma
^\alpha|k_n|^\alpha|_{n=1})/(\tau_1+\tau_2)$. In this approximation we replace the sum $L^{-1}
\sum_{n=-\infty}^\infty$ by the integral $\int_{-\infty}^\infty d k_n/(2\pi)$,
obtaining
\begin{equation}
\label{eq:W0levy}
W_0(u)\sim\frac{1}{2\sqrt{D\tau_1^{-1}}}+\frac{\tau_1/[\alpha\sin\left(\pi/
\alpha\right)\sigma]}{ \left[u(\tau_1+\tau_2)\right]^{1-1/\alpha}}.
\end{equation}
For shorter times (corresponding to larger $u$) we discard the
subdominant second term in Eq.~(\ref{eq:W0levy}). Laplace inversion of
Eq.~(\ref{eq:pfaW0}) then produces
\begin{equation}
p_{\rm fa}(t)\sim \frac{2\sqrt{D\tau_1}}{L (\tau_1+\tau_2)}.
\label{eq:pfa1}
\end{equation}
At later times  (smaller $u$) the second term in Eq.~(\ref{eq:W0levy})
dominates, and the plateau (\ref{eq:pfa1}) turns into
\begin{equation}
p_{\rm fa}(t)\sim \frac{\alpha}{2}\left[\sin\left(\frac{\pi}{\alpha}\right)\right]^2
\frac{v\tau_2}{L\left(\tau_1+\tau_2\right)^{1/\alpha}t^{1-1/\alpha}}.
\label{eq:pfa2}
\end{equation}
The crossover between these two regimes occurs when the values of expressions
(\ref{eq:pfa2}) and (\ref{eq:pfa1}) become equal, i.e., at
\begin{equation}
t \sim (\tau_1+\tau_2)\left\{\frac{\alpha[\sin(\pi/\alpha)]^2v\tau_2}{4\sqrt{D
\tau_1}}\right\}^{\alpha/(\alpha-1)}.
\end{equation}

\begin{figure}
\includegraphics[width=8cm]{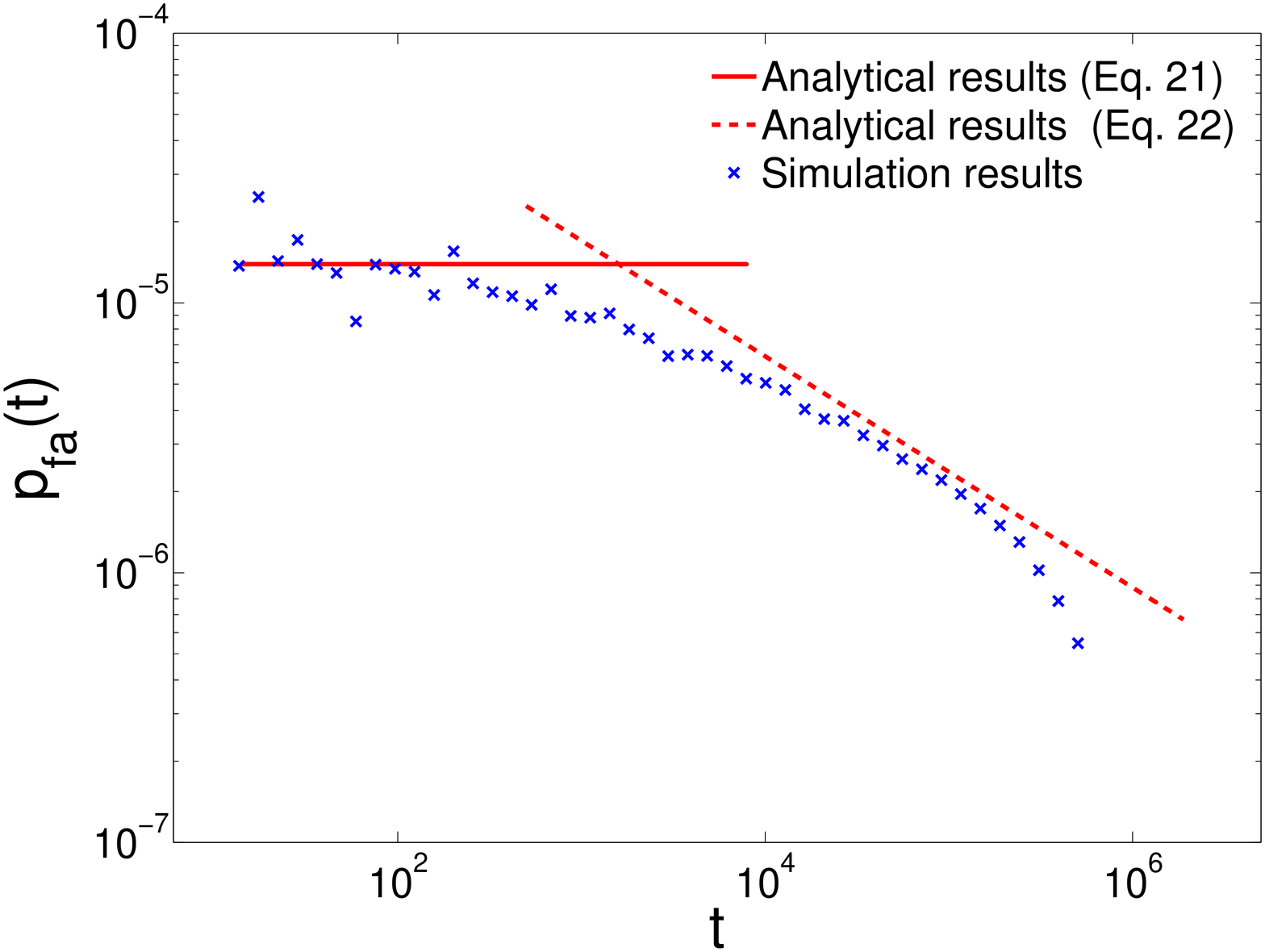}
\caption{First arrival density versus time. The crosses are simulation data,
while the straight lines are the intermediate regimes of Eq. (\ref{eq:pfa1})
and Eq. (\ref{eq:pfa2}). Parameters are: $\tau_1=35$, $\tau_2=50$,
$L=10^4$, $\alpha=1.75$, $v=1$ and $D=1$.}
\label{arrival}
\end{figure}

Note that in Eq.~(\ref{eq:pfa1}), $2\sqrt{D\tau_1}$ is the average length
scanned in a search event. Division by $L$ yields the probability to find the
target during this phase, and $1/(\tau_1+\tau_2)$ is the rate at which the
search phase itself occurs. A crucial part in this interpretation is that the
probability of searching in a previously scanned area is negligible. This
assumption will break down at some point because
of the searcher's lack of orientational memory. The searcher will
then begin to revisit explored regions with a reduced probability of
finding the target as a result. This causes the crossover
to the power-law behavior (\ref{eq:pfa2}). Fig.~\ref{arrival} shows the
turnover from plateau to inverse power-law of the first arrival. At even
longer times, finite size effects cause a turnover to an exponential
decay.

From Eq.~(\ref{eq:pfa2}) the advantage of having $\alpha$ close to unity at
large $L$ becomes evident: the presence of rare but long relocation events
reduces the risk of rescanning already visited areas which will be important
for large $L$. However, the downside to choosing an $\alpha$-value too close
to 1 is that an increased amount of very long relocations implies an increased
amount of very short ones too, as the average distance is fixed by $v\tau_2$
\cite{REMM}. This means that the crossover to the
less favorable situation described by Eq. (\ref{eq:pfa2}) happens earlier,
so that larger $\alpha$ becomes more efficient for shorter search
times relevant at smaller $L$.

From a more general perspective, intermittent strategies are beneficial when
purely diffusive search would slow down over time due to the increasing number
of returns to previously scanned areas (oversampling). Choosing an exponential
strategy for the relocation events, however, only partially solves this problem,
as this strategy is still governed by the central limit theorem (CLT). Thus, the
problem of oversampling merely becomes postponed to later times. Conversely,
L{\'e}vy-intermittent strategies are not bound to the CLT, rendering them
advantageous in the search for rare targets. Although less pronounced, the
problem of oversampling still occurs in two dimensional search
studied in \cite{benichou06b}. Thus, L{\'e}vy strategies are expected to be
advantageous in this case, as well.

We have shown that for intermittent search strategies
L{\'e}vy distributed relocations are advantageous over exponential
distributions when targets are sparse, because rare long relocations reduce
the eventually occurring problem of oversampling. Thus we advocate that
intermittent strategies should not be thought of as alternatives to
L{\'e}vy strategies, as suggested in Ref.~\cite{benichou06b}. In contrast,
the combination of intermittent search and L{\'e}vy relocation strategies
turns out to be beneficial.

Our analysis relies on the assumption that each relocation is pointed
toward a random direction. This will be a good model for ``non-intelligent''
search, similar to bacterial movement in absence of chemical or
temperature gradients during which tumbling motion changes with directed
motion \cite{chemo}. Intelligent creatures will improve the target
search by partial or complete memory, avoiding previously visited locations.
It will be interesting to study in more detail models with search memory.


\begin{thebibliography}{99}
\bibitem{REM} Note the typo in the expression for $\langle t\rangle$ in Eq.~(5)
of Ref.~\cite{benichou06}: $\coth(1/[2\alpha L])$ should be $\coth(\alpha L/2)$.
\bibitem{REMM} Corresponding to the fact that the central part close to $x=0$
of L{\'e}vy stable laws with a smaller index $\alpha$ is more pronounced
than for larger $\alpha$ or a Gaussian.
\end{thebibliography}
\end{document}